\documentclass{cicahep2015}
\usepackage{graphicx,amssymb,amsmath,times}
\begin{document}
\title[The pitch angle paradox and radiative life times in a synchrotron source]
{The pitch angle paradox and radiative life times in a synchrotron source}
\author[Ashok K. Singal]{Ashok K. Singal\\
Astronomy and Astrophysics Division, Physical Research Laboratory, 
Navrangpura, Ahmedabad - 380 009, India}
\presenter{(asingal@prl.res.in); A\&A7, Oral, CICAHEP15.169}
\talktype{PS2}
\pagerange{1--6}
\maketitle
\begin{abstract}
In synchrotron radiation there is a paradox whether or not the pitch angle of a radiating charge varies. 
The conventional wisdom is that the pitch angle does not change during the radiation process. 
The argument is based on Larmor's radiation formula, where in a synchrotron case 
the radiation power is along the instantaneous direction of motion of the charge. Then the  
momentum loss will also be parallel to that direction and therefore the pitch angle of the charge would 
remain unaffected. The accordingly derived formulas for 
energy losses of synchrotron electrons in radio galaxies are the standard text-book 
material for the last 50 years. However, if we use the momentum transformation laws from special relativity, 
then we find that the pitch angle of a radiating charge varies. While the velocity 
component parallel to the magnetic field remains unaffected, the perpendicular component does 
reduce in magnitude due to radiative losses, implying a change in the pitch angle. This apparent paradox 
is resolved when effects on the charge motion are calculated not from Larmor's formula but from 
Lorentz's radiation reaction formula. We derive the exact formulation by taking into account the change of 
the pitch angle due to radiative losses. From this we first time derive the characteristic decay time of 
synchrotron electrons over which they turn from highly relativistic into mildly relativistic ones. 
\end{abstract}
\section{INTRODUCTION}
Synchrotron radiation is of extreme importance in many relativistic plasma and is  
widely prevalent in extragalactic radio sources, supernovae remnants, the Galaxy, 
and many other astrophysical phenomena. A power-law 
spectrum is the main characteristic of this radiation process. As electrons 
emit radiation and thereby lose energy, the radiation spectrum steepens. This 
steepening of the spectrum causes even a break in the spectrum slope. This break 
is a direct indication of the radiative life-time of electrons and tells 
us about the age of the source of synchrotron radiation. Therefore it is important 
to understand these radiation losses in detail.

Formulas for synchrotron radiative losses were derived more than about 
50 years back and have been in use ever since for calculating radiative 
losses in a variety of radio sources. In these formulas, it has always been assumed that the pitch angle of the radiating 
charged particle remains constant and the 
dynamics and the life-time of radiating electrons are accordingly derived \cite{Kardashev}. 
This formulation is now a standard text-book material  \cite{Pacholczyk, Rybicki}.
However, it turns out that this formulation is not relativistically covariant. It will be 
shown here that in the case of synchrotron losses, the pitch angle in general changes. 
We shall derive the exact formulation for radiative losses, taking into account the pitch angle changes. 
\section{SYNCHROTRON POWER LOSS}
We use Gaussian (cgs) system of units throughout. 
A relativistic charged particle, say, an electron of charge $-e$, rest mass $m_{0}$, 
having a velocity $\mbox{\boldmath $\beta$}={\bf V}/c$ and energy ${\cal E}=m_{0}c^2\gamma$ (with Lorentz factor 
$\gamma=1/\surd(1-\beta ^{2})$), moves in a magnetic field $B$
in a helical path with $\theta$ as the pitch angle, defined  as the angle that the velocity 
vector makes with the magnetic field direction. We assume the magnetic field to be 
uniform, say, along the z-axis (Fig.(1)). As there is no force component due to the magnetic field parallel to itself, 
a charge with a velocity component 
$\beta_{\parallel}$ only along the $z$-axis keeps on moving unaffected by the field. 
\begin{figure}[ht]
\begin{center}
\includegraphics[scale=0.45]{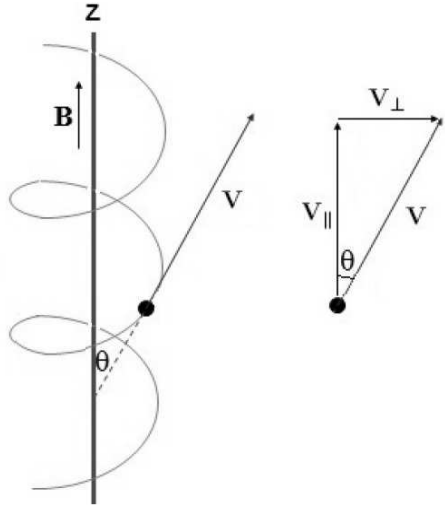}
\caption{Helical motion of the electron moving with a velocity ${\bf V}$ along pitch angle 
$\theta$ in a uniform magnetic field ${\bf B}$. ${\bf V}_{\parallel}$ is the velocity component parallel to 
${\bf B}$ while ${\bf V}_{\perp}$ is the perpendicular velocity component.} 
\end{center}
\end{figure}

The charge spiraling in a magnetic field radiates in the forward direction of its instantaneous motion. 
For a highly relativistic motion of the charge, like in a synchrotron case, all 
the radiated power, as calculated from Larmor's formula (or rather from Li\'{e}nard's formula), 
is confined to a narrow cone of angle $1/\gamma$ around the instantaneous direction of motion 
of the charge. Therefore the momentum carried by the radiation will be along the direction of motion of the charge. 
From the conservation of momentum it can then be construed that the radiation should 
cause only a decrease in the magnitude of the velocity vector 
without affecting its direction. As a result it is expected 
that the pitch angle $\theta$ of the motion should not change due to radiative losses \cite{Kardashev}.
Thus the ratio $V_\perp/V_\parallel=\beta_\perp/\beta_\parallel = \tan\theta$, will not change. 

With the pitch angle as a constant of motion, half life-times of radiating electrons have been calculated, 
using an approximate power loss formula \cite{Kardashev, Pacholczyk, Rybicki},
\begin{equation}
\label{eq:30.1}
\frac{{\rm d}{\cal E}}{{\rm d}t} = -\zeta \sin^2 \theta\; {\cal E}^2\;,
\end{equation}
where $\zeta = {2e^4\, B^2}/ ({3m_{o}^4c^7})=2.37 \times 10^{-3}B^2  {\rm erg}^{-1} {\rm s}^{-1}$.
Let ${\cal E}_{\rm 0}$ be the initial energy  at $t=0$, then from Eq.~(\ref{eq:30.1}) the energy of the radiating charge at $t=\tau$ is 
calculated to be,
\begin{equation}
\label{eq:30.4}
{\cal E}= \frac{{\cal E}_{\rm 0}}{1+\zeta\sin^2 \theta\;  \tau\; {\cal E}_{\rm 0}}\;.
\end{equation}
From Eq.~(\ref{eq:30.4}) it follows that the electron loses half of its energy in a time 
$\tau_{1/2} = 1/(\zeta\sin^2 \theta \;{\cal E}_{\rm o})$.  

One thing we note from Eq.~(\ref{eq:30.4}) is that it can be true only 
for a highly relativistic charge ($\gamma={\cal E}/(m_{0}c^2)\gg1$) and that it is not a general equation. 
This can be seen immediately from $\tau\rightarrow\infty$,   
where ${\cal E}\rightarrow 0$ implying $\gamma \rightarrow 0$, while we know that $\gamma\ge1$ always. Actually an 
approximation $\beta \approx 1$ has been used right from the beginning and instead of Eq.~(\ref{eq:30.1}),
the exact equation for the energy loss rate is \cite{Jackson, Melrose, Longair}, 
\begin{eqnarray}
\label{eq:30.1b}
\frac{{\rm d}{\cal E}}{{\rm d}t} = -\zeta \beta ^2\sin^2 \theta\;{\cal E}^2.
\end{eqnarray}
We can rewrite the power loss rate in terms of the Lorentz factor 
$\gamma$ ($={\cal E}/(m_{0}c^2)$, which implies expressing energy in units of the rest mass energy), to write,
\begin{equation}
\label{eq:30.2}
\frac{{\rm d}{\gamma}}{{\rm d}t} = -{\eta}\beta ^2\sin^2 \theta\;\gamma^2.
\end{equation}
Here ${\eta}=2e^4 B^2/(3m_{\rm o}^3c^5)=1.94 \times 10^{-9}B^2\; {\rm s}^{-1}$.
\section{THE PITCH ANGLE PARADOX}
Consider a charge particle in its gyrocenter (GC) frame ${\cal K}'$, which is moving with a velocity $\beta_{\parallel}$ with 
respect to the lab frame ${\cal K}$. In ${\cal K}'$ therefore 
the charge has no component of velocity parallel to the magnetic field and has 
only a circular motion in a plane perpendicular to the magnetic field (with a pitch angle $\theta=\pi/2$). 
In this frame, due to radiative losses by the charge, there will be a decrease in the velocity which is solely  
in a plane perpendicular to the magnetic field, and the charge will never ever attain a velocity component parallel to the magnetic field.

Now we look at this particle from the lab frame ${\cal K}$, in which the charge has (at least to begin with) a motion, 
$\beta_{\parallel}$ along 
the magnetic field. Since in the inertial frame ${\cal K}'$, the charge never gets a velocity component parallel 
to the magnetic field  and the two inertial frames (${\cal K}$ and ${\cal K}'$) continue to move with reference to each other with a 
constant $\beta_\parallel$, the parallel component of velocity of the charge should remain unchanged even in ${\cal K}$. However, magnitude 
of the perpendicular component of velocity is continuously decreasing because of radiative losses, therefore the pitch angle of the charge, 
$\theta = \tan^{-1} (\beta_\perp/\beta_\parallel)$, should decrease continuously with time and 
the velocity vector of the charge should increasingly align with the magnetic field vector.

Thus we have a paradox here. While conservation of momentum argument led us to the conclusion that 
the pitch angle of the charge is a constant, the second argument from relativistic transformation 
considerations showed that the pitch angle will be progressively reducing as the charge radiates 
with time. Which of the two is true then? It turns out that the second argument is correct and we shall show that 
the pitch angles of the radiating charges decrease continuously and due to that their angular distribution in the 
momentum space changes with time. Even if to begin with there were an isotropic energy distribution of electrons in a synchrotron source, 
electrons radiating synchrotron radiation develop a pitch angle anisotropy because of $\sin^2\theta$ dependence of the radiated power 
(Eq.~(\ref{eq:30.2})). Then there is the additional fact that the pitch angle of radiating electrons monotonically decreases with time, 
and as we shall show the rate of change of the pitch angle depends upon the value of the pitch angle itself. 

Instead of calculating the effects of radiation on charge energy from Larmor's formula (or its relativistic generalization 
Li\'{e}nard's formula) if we use the Lorentz's radiation reaction formula and apply it in frame ${\cal K}'$, 
we get a force along $\ddot{\mbox{\boldmath $\beta$}'}$, 
which is in a direction opposite to the velocity vector in ${\cal K}'$, and the charge 
accordingly would have a deceleration vector only in a plane perpendicular to the magnetic field. It should be noted that 
in all reference frames gyro acceleration $\dot{\mbox{\boldmath $\beta$}}$ 
always lies in the plane perpendicular to the magnetic field and so is the vector $\ddot{\mbox{\boldmath $\beta$}}$ therefore.
And a relativistic transformation of acceleration due to radiation losses, between frames ${\cal K}$ to ${\cal K}'$, 
gives a non-zero vector only along the direction perpendicular to the magnetic field  and a nil acceleration along the parallel 
direction \cite{Tolman}, consistent with conclusions reached based on the theory of relativity. 

There is another way of arriving at the paradox. Larmor's formula says that a charge moving with non-relativistic speeds 
radiates energy at a rate $\propto \dot{\beta}^2$. However, the radiation pattern of such a charge has a $\sin^2\phi$ dependence 
\cite{Jackson, Panofsky, Griffith}, about the direction of acceleration. Due to this azimuthal symmetry 
the net momentum carried by the radiation is nil. Therefore the charge too cannot be losing momentum. Thus we have the paradox of a 
radiating charge losing its kinetic energy but without a corresponding change in its momentum. 
\section{SYNCHROTRON LOSSES AND THE RADIATIVE LIFE TIMES}
We first calculate the power losses in the GC frame ${\cal K}'$, 
where pitch angle is a constant ($\theta=\pi/2$) and  thus the standard formulation should be applicable,  
and then using special relativistic transformations, convert them to the lab frame ${\cal K}$. 

In the GC frame ${\cal K}'$, there is no motion along the $z$ direction 
and the charge moves in a circle in the $x$-$y$ plane. The velocity $\beta'$ of the charge as well as the force $F'$ due to radiation losses 
are perpendicular to the $z'$-axis in frame ${\cal K}'$ and there is hardly any ambiguity about that. 
From the 4-force transformation \cite{Tolman, Mould} to the lab frame ${\cal K}$, with respect to which the GC frame ${\cal K}'$ 
is moving along the $z$ direction with a velocity $\beta_\parallel$, we get,
\begin{equation}
\label{eq:10.1.6}
{F}_\parallel=F' \gamma_\parallel \beta_\parallel {\beta}_\perp\;,
\end{equation}
\begin{equation}
\label{eq:10.1.7}
{F}_\perp=\frac{F'}{\gamma_\parallel}\;.
\end{equation}

There is of course no acceleration component $\dot{\beta}_\parallel$ along the $z$-axis, even though a finite 
parallel force component ${F}_\parallel$ exists. From a relativistic transformation of acceleration 
\cite{Tolman} we can verify that there is no parallel component of acceleration in frame ${\cal K}$ if it is zero in frame ${\cal K}'$ 
(i.e.,  $\dot{\beta}_{\parallel}=0$ if $\dot{\beta}'_{\parallel}=0$). Actually in frame ${\cal K}$, a force component along $z$ direction 
shows up solely because of a rate of change of $\gamma$ due to $\dot{\beta}_{\perp}$, even though $\dot{\beta}_{\parallel}=0$.
It can be recalled that in special relativity, force and acceleration vectors are not
always parallel, e.g., in a case where force is not parallel to the velocity vector, the acceleration need not be 
along the direction of the force. When the applied force is either parallel to or perpendicular to the velocity vector, it is 
only then that the acceleration is along the direction of force \cite{Tolman}. 
It has to be further kept in mind that the acceleration we are talking 
about here is not that due to the force by the magnetic field on the moving charge (which is perpendicular to the 
instantaneously velocity of the charge), but the acceleration (or rather a deceleration) caused on the charge due to the 
radiation reaction force. An alternative derivation of radiation losses and the pitch angle changes based on radiation reaction force is  
available in Singal \cite {Singal16}.

In Eq.~(\ref{eq:30.2}) $\sin \theta$ is a variable, but we can write this equation for the GC 
frame ${\cal K}'$, where pitch angle is always a constant ($\theta'=\pi/2$). Then we have, 
\begin{equation}
\label{eq:30.7}
\frac{{\rm d}{\gamma}'}{{\rm d}t'} =  -{\eta}\beta'^2\gamma'^2= -{\eta}(\gamma'^2-1)\;.
\end{equation}
 
Equation~(\ref{eq:30.7}) has a solution,
\begin{equation}
\label{eq:30.2c1}
\tanh^{-1}\frac{1}{\gamma'}= {\eta} t' + a\;.
\end{equation}
Let ${\gamma'_{\rm o}}$ be the initial energy at $t'=0$ in frame ${\cal K}'$, then 
$1/{\gamma'_{\rm o}}=\tanh(a)$ and at time $t'=\tau'$ we have,
\begin{equation}
\label{eq:30.2e3}
\tanh^{-1}\frac{1}{\gamma'} = \tanh^{-1}\frac{1}{{\gamma'_{\rm o}}}+{\eta} \tau'\;.
\end{equation}
which complies with the expectations that as $\tau' \rightarrow \infty$, $\gamma' \rightarrow 1$.

Now a transformation between ${\cal K}'$ and ${\cal K}$ gives 
$\gamma' \gamma_\parallel  = \gamma$ and  $\gamma' \beta' = \gamma \beta_\perp$ or $\beta' = \gamma_\parallel \beta_\perp$ \cite{Melrose}. 
Also we have ${\rm d}t/{\rm d}t'=\gamma_\parallel$ or $\tau = \tau' \gamma_\parallel$. 
For the transformation of acceleration we then get $\dot{\beta}'  = \gamma_\parallel^2 \dot{\beta}_\perp$  
with $\dot{\beta}_\parallel = \dot{\beta}'_\parallel/\gamma_\parallel^3 = 0$. 

Equation~(\ref{eq:30.2e3}) can then be transformed in terms of quantities expresses in the lab frame ${\cal K}$,
\begin{equation}
\label{eq:30.2e4}
\tanh^{-1}\frac{\gamma_\parallel}{\gamma} = \tanh^{-1}\frac{\gamma_\parallel}{{\gamma_{\rm o}}}+\frac{{\eta} \tau}{\gamma_\parallel}\;.
\end{equation}
This is a general solution for all values of $\gamma$. 
We can rewrite it as,
\begin{equation}
\label{eq:30.2e2}
\gamma = \gamma_\parallel\; \frac{{\gamma_{\rm o}}+\gamma_\parallel\tanh({\eta} \tau/\gamma_\parallel)}
{{\gamma_{\rm o}}\tanh({\eta} \tau/\gamma_\parallel)+\gamma_\parallel}\;.
\end{equation}
For an initially ultra relativistic charge ($\gamma_0\gg1, \beta_0 \approx 1$), we have 
$1/\gamma_\parallel =\surd(1-\beta_0^{2}\cos^2 \theta_0)\approx \sin\theta_0$. That also implies that 
(except for initially small pitch angle cases) $\sin \theta_0 \gg 1/\gamma_0$ or $\gamma_\parallel / \gamma_0 \ll 1$, and from 
Eq.~(\ref{eq:30.2e4}) we could write,
\begin{equation}
\label{eq:30.2e1a}
\tanh^{-1}\frac{1}{\gamma \sin\theta_0} = {{\eta} \tau}{\sin\theta_0}\;.
\end{equation}
This implies that for $\tau = \gamma_\parallel/ \eta \approx 1/ (\eta \sin \theta_0)$, we have $\gamma \approx 1.3/\sin \theta_0$. 
Thus even if an electron had started with an almost infinite energy, it loses most of its kinetic energy in a time interval of the order 
of $1/ \eta$, reducing to perhaps a mildly relativistic status (for not too small an initial pitch angle). 
For instance let us consider $\gamma_0=10^3$ and $\theta_0=\pi/4$, then $\gamma_\parallel \approx 1/\sin \theta_0=\surd{2}$, then from 
Eq.~(\ref{eq:30.2e4}) or Eq.~(\ref{eq:30.2e1a}) we get for $\tau = 1/\eta$, $\gamma=2.3$. In another example, taking $\gamma_0=10^4$ and 
$\theta_0=\pi/3$,  for $\tau = 1/\eta$ we get $\gamma_\parallel \approx 1/\sin \theta_0=2/\surd{3}$ and $\gamma=1.7$.
Thus  $1/ \eta$ represents the characteristic decay time of synchrotron electrons over 
which they turn from ultra relativistic into mildly relativistic ones. 
\section{REDUCTION IN THE PITCH ANGLE}
Equation~(\ref{eq:30.7}) can be also written as,
\begin{equation}
\label{eq:30.8a}
\gamma'^3 \dot{\beta}' {\beta'} =  -{\eta} \beta'^2 {\gamma}'^2,
\end{equation}
or
\begin{equation}
\label{eq:30.8b}
\frac{\dot{\beta}'}{\beta'} =  \frac{-{\eta}} {\gamma'}\;.
\end{equation}

Then transforming to the lab frame ${\cal K}$ we have,
\begin{equation}
\label{eq:30.8c}
\frac{\dot{\beta}_\perp}{\beta_\perp} =  \frac{-{\eta}} {\gamma} \;, 
\end{equation}
Both $\beta$ and $\theta$ in $\beta_\perp=\beta \sin \theta$ are functions of time. Therefore we can rewrite Eq.~(\ref{eq:30.8c}) as,
\begin{equation}
\label{eq:30.6c1}
{\beta \cos \theta \frac{{\rm d}\theta}{{\rm d}t} + \dot{\beta} \sin \theta} =  \frac{-{\eta}\beta \sin \theta} {\gamma}\;.
\end{equation}
Also from $\dot{\beta}_\parallel=0$ we get,
\begin{equation}
\label{eq:30.6d}
\beta \sin \theta \frac{{\rm d}\theta}{{\rm d}t} = \dot{\beta} \cos \theta\;.
\end{equation}
Eliminating $\dot{\beta}$ from Eqs.~(\ref{eq:30.6c1}) and (\ref{eq:30.6d}), we get,
\begin{equation}
\label{eq:30.6e}
\frac{{\rm d}\theta}{{\rm d}t}= \frac{-{\eta}\sin \theta \cos \theta}{\gamma}=\frac{-{\eta}\sin 2\theta}{2\gamma}\;.
\end{equation}
This is the relation for the rate of change of the pitch angle of a charge undergoing synchrotron radiative losses. 
The negative sign implies that the pitch angle decreases with time and the velocity vector gets increasingly aligned with the magnetic field. 
The rate of alignment is very slow for low pitch angles ($\theta \approx 0$) as well as for high pitch angles ($\theta \approx \pi/2$), 
and the highest rate of change of the pitch angle is for $\theta=\pi/4$.

With the help of Eq.~(\ref{eq:30.2e4}), we can integrate Eq.~(\ref{eq:30.6e}),
\begin{equation}
\label{eq:30.6h}
\int^{\theta}_{\theta_0}\frac{{\rm d}\theta}{\sin \theta \cos \theta}
= -\int^{\tau}_{0}\frac{{\eta}{\rm d}t}{\gamma_\parallel} \tanh \left(\frac{\eta t}{\gamma_\parallel} + a\right)\;,
\end{equation}
where $a=\tanh^{-1}({\gamma_\parallel/\gamma_{\rm o}})$. This gives us,
\begin{equation}
\label{eq:30.6h}
\ln\frac{\tan \theta} {\tan \theta_0}= \ln\frac{\cosh (a)}{\cosh \left(\frac{{\eta} \tau}{\gamma_\parallel} + a \right)}\;.
\end{equation}
or  
\begin{equation}
\label{eq:30.6i}
{\tan \theta} = \frac{{\tan \theta_0}}{\cosh \left(\frac{{\eta} \tau}{\gamma_\parallel}\right)
+ \frac{\gamma_\parallel}{\gamma_0} \sinh \left(\frac{{\eta} \tau}{\gamma_\parallel} \right)}\;.
\end{equation}

In Eq.~(\ref{eq:30.6i}), $\theta < \theta_0$, because pitch angle always reduces with time. There are many notable points. If 
$\theta_0 =\pi/2$, then $\theta =\pi/2$ also, which is because if the pitch angle is $\pi/2$, then the radiating electron always moves in 
a circular path in the plane perpendicular to the magnetic field. And if $\theta_0 =0$, then $\theta=0$ too as there is no more reduction 
in the pitch angle. For any $0<\theta_0 <\pi/2$, $\theta \rightarrow 0$ as $\tau \rightarrow \infty$. For large $\gamma_0$ values, 
\begin{equation}
\label{eq:30.6j}
{\tan \theta} = \frac{{\tan \theta_0}}{\cosh \left({{\eta} \tau}{\sin \theta_0}\right)}\;,
\end{equation}
which can be used to estimate change in pitch angle with time. For example for say, $\theta_0 =\pi/3$, and $\theta =\pi/6$, 
$\cosh ({\eta} \tau \sin \theta_0)=3$, which gives $\tau \approx 2 /\eta$ for this change in the pitch angle.
Thus there are appreciable pitch angle changes in time $\tau \sim 1/\eta$ (except for 
in the vicinity of very small pitch angles). 

All charges of a given energy and pitch angles, directed towards the observer in a narrow angle $1/\gamma$ around the line of sight 
not only lose energy but will also get shifted outside the angle $1/\gamma$ around the line of sight towards the observer, in a time 
$\tau \sim 1/\eta$. Thus in a mono-energetic and a narrow pitch angle distribution, the pitch angle changes might be quite relevant. 
But it may be of less importance when there is a wide angular (isotropic!) distribution of pitch angles.


\begin{thebibliography}{99}
\bibitem{Kardashev}N. S. Kardashev, Sov. Astr. - AJ, 6, 317 (1962)
\bibitem{Pacholczyk} A. G. Pacholczyk, Radio astrophysics, Freeman, San Francisco (1970)
\bibitem{Rybicki}G. B. Rybicki \& A. P. Lightman, Radiative processes in astrophysics, Wiley, New York (1979)
\bibitem{Jackson} J. D. Jackson, Classical electrodynamics, 2nd ed., Wiley, New York (1975)
\bibitem{Melrose} D. B. Melrose, ApL, 8, 35 (1971)
\bibitem{Longair} M. S. Longair, High energy astrophysics, Cambridge Univ. Press, Cambridge (2011)
\bibitem{Panofsky} W. K. H. Panofsky,  M. Phillips, Classical electricity and magnetism, 2nd ed., Addison-Wesley, Massachusetts (1962)
\bibitem{Griffith} D. J. Griffiths, Introduction to electrodynamics, 3rd ed., Prentice, New Jersey (1999)
\bibitem{Tolman} R. C. Tolman, Relativity thermodynamics and cosmology, Clarendon, Oxford (1934) 
\bibitem{Mould} R. A. Mould, Basic relativity, Springer, New York (1994)
\bibitem{Singal16} A. K. Singal, Mon. Not. R. astr. Soc., DOI: 10.1093/mnras/stw349 (2016) 
\end{thebibliography}
\end{document}